\documentclass{ws-procs10x7}
\usepackage{amsmath}
\usepackage{amssymb}
\usepackage{euscript}
\usepackage{fancybox}
\usepackage{color}

\def\mathswitchr#1{\relax\ifmmode{\mathrm{#1}}\else$\mathrm{#1}$\fi}

\newcommand {\pslash}{\hbox{$\not\hbox{\kern-2.3pt $p$}$}}

\usepackage{cite}

\usepackage{epic}

\def\alf1{ {\alpha\over\pi} }

\begin{document}
 
\title{Resummed Quantum Gravity}
\author{B.F.L. Ward}

\address{Department of Physics,\\
Baylor University, Waco, Texas, USA}

\twocolumn[\maketitle\abstract{
We present the current status of the a new approach to quantum general relativity based on the exact resummation of its perturbative series
as that series was formulated by Feynman. We show that the
resummed theory is UV finite and we present some phenomenological
applications as well.\\
\vspace{0.5cm}
\centerline{BU-HEPP-06-08, Oct., 2006, {\it presented at ICHEP06}}
}]
%


\def\Kmax{K_{\rm max}}\def\ieps{{i\epsilon}}\def\rQCD{{\rm QCD}}
\renewcommand{\theequation}{\arabic{equation}}
\font\fortssbx=cmssbx10 scaled \magstep2
\renewcommand\thepage{}
\parskip.1truein\parindent=20pt\pagenumbering{arabic}\par
\section{\bf Introduction}\label{intro}\par
The successful classical generalization of Newton's law of gravity
by Einstein in the
general theory of relativity is one of the outstanding achievements
of 20th century physics. Left over for solution in the 21st century is 
the problem of the union of general relativity and quantum mechanics.
While the superstring theory~\cite{gsw,jp} is currently the only
accepted solution of this problem with no unresolved theoretical issues,
the lack of any experimental verification of superstring theory
invites consideration of other approaches. Indeed, the loop-quantum gravity
approach is yet another possible solution~\cite{lpqg1}, though it may still
have some issues of principle. Accordingly, in Refs.~\cite{BW1,BW2,BW3,BW4},
we have introduced a new approach to this outstanding problem
of quantum general relativity. 
\par
Our approach is based on methods 
well-tested~\cite{yfs,yfs1}
in the theory of higher-order radiative corrections in high precision
studies at LEP1 and LEP2 and recently extended to high precision LHC
physics scenarios~\cite{qcdyfs,qced}. We have called the new theory
resummed quantum gravity, as we follow Feynman's formulation~\cite{ff1,ff2} 
of Einstein's
theory as a point particle quantum field theory and, taking a hint from the work of Yennie, Frautschi and Suura~\cite{yfs}
in which they discovered that resumming the infrared effects in the electron
propagator in QED leads to an improved convergence in the UV for the
QED loop corrections that involved that propagator, we show that
resumming the large infrared effects in quantum general relativity leads 
in fact to a UV finite result. This then is a pure union of the ideas of
Bohr and Einstein.\par
The discussion proceeds as follows. After reviewing Feynman's formulation of
Einstein's theory in the next section, we show how resummation
renders the theory UV finite in Section 3. Section 4 then presents
some phenomenology of the new theory. Section 5 sums up the discussion.\par
\section{Einstein's Theory as Formulated by Feynman}
The basic idea of Feynman~\cite{ff1,ff2} is that quantum general relativity
is a point particle field theory where the graviton represents quantum
fluctuations about the background metric $\eta_{\mu\nu}$ 
of space-time: $g_{\mu\nu}=\eta_{\mu\nu}+2\kappa h_{\mu\nu}$, where $\kappa=\sqrt{8\pi G_N}$ with $G_N$ equal to
Newton's constant so that $G_N=1/M_{Pl}^2$ where $M_{Pl}=1.22\times 10^{19}$GeV
is the Planck mass. This means that, after we specialize the Lagrangian
of the world to the scalar Higgs sector for definiteness ( if we show that the
Higgs-graviton system is UV finite, the inclusion of the spinning particles
will follow without essential complication), then we are led by Feynman~\cite{ff1,ff2}
to consider the Lagrangian
{\small
\begin{equation}
\begin{split}
{\cal L}(x) &= -\frac{\sqrt{-g}}{2\kappa^2} R
            + \frac{\sqrt{-g}}{2}\left(g^{\mu\nu}\partial_\mu\varphi\partial_\nu\varphi - m_o^2\varphi^2\right)\\
            &= \frac{1}{2}{\big\{} h^{\mu\nu,\lambda}\bar h_{\mu\nu,\lambda} - 2\eta^{\mu\mu'}\eta^{\lambda\lambda'}
\bar{h}_{\mu_\lambda,\lambda'}\eta^{\sigma\sigma'}\\
&\bar{h}_{\mu'\sigma,\sigma'}{\big\}}
          + \frac{1}{2}{\big\{}\varphi_{,\mu}\varphi^{,\mu}-m_o^2\varphi^2 {\big\}} \\
&-\kappa {h}^{\mu\nu}{\big[}\overline{\varphi_{,\mu}\varphi_{,\nu}}+\frac{1}{2}m_o^2\varphi^2\eta_{\mu\nu}{\big{]}}\\
            & \quad - \kappa^2 [ \frac{1}{2}h_{\lambda\rho}\bar{h}^{\rho\lambda}{\big{(}} \varphi_{,\mu}\varphi^{,\mu} - m_o^2\varphi^2 {\big{)}} \\
&- 2\eta_{\rho\rho'}h^{\mu\rho}\bar{h}^{\rho'\nu}\varphi_{,\mu}\varphi_{,\nu}] + \cdots \\
\end{split}  
\label{eq1}
\end{equation}}\noindent
where $\varphi_{,\mu}\equiv \partial_\mu\varphi$ and 
we have the metric
$g_{\mu\nu}(x)=\eta_{\mu\nu}+2\kappa h_{\mu\nu}(x)$ with
$\eta_{\mu\nu}={\text diag}\{1,-1,-1,-1\}$ and 
$\bar y_{\mu\nu}\equiv \frac{1}{2}\left(y_{\mu\nu}+y_{\nu\mu}-\eta_{\mu\nu}{y_\rho}^\rho\right)$ for any tensor $y_{\mu\nu}$.
Feynman has thus formulated Einstein's theory as 
just another point particle field theory 
for which he has already worked-out the Feynman rules in Ref.~\cite{ff1,ff2}.
\par 

\section{Resummed Quantum Gravity}
In Refs.~\cite{BW1,BW2,BW3,BW4}, we have extended the approach of
Yennie, Frautschi and Suura~\cite{yfs} to the Lagrangian in (\ref{eq1})
by re-arranging the Feynman series for the 1PI 2-point functions, exactly.\par
\begin{figure}
\begin{center}
\epsfig{file=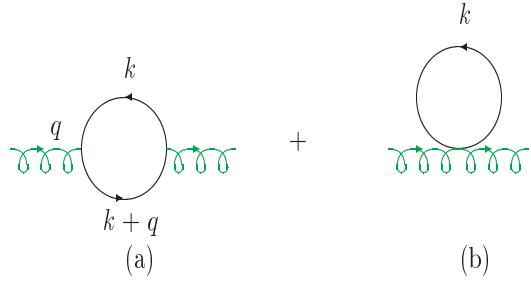,width=77mm,height=38mm}
\end{center}
\caption{\baselineskip=7mm  The scalar one-loop contribution to the
graviton propagator. $q$ is the 4-momentum of the graviton.}
\label{fig1}
\end{figure}

Specifically, as we show in Ref.~\cite{BW1,BW4}, our exact re-arrangement of the
Feynman series for the scalar 1PI 2-point function leads to the result
\begin{equation}
i\Delta'_F(p)|_{resummed} = \frac{ie^{B''_g(p)}}{(p^2-m^2-\Sigma'_s(p)+i\epsilon)}
\label{indn9}
\end{equation}
where 
\begin{equation}
\Sigma'_s(p)\equiv \sum_{\ell=1}^{\infty}\Sigma'_\ell(p).
\label{indn10}
\end{equation}
when $\Sigma'_\ell(p)$ is the corresponding $\ell$-loop 1PI 2-point function
residual
and where the exponent $B''_g(p)$ is given by
{\small ~~~($\Delta =k^2 - m^2$)
\begin{equation}
\begin{split} 
B''_g(k)&= -2i\kappa^2k^4\frac{\int d^4\ell}{16\pi^4}\frac{1}{\ell^2-\lambda^2+i\epsilon}\\
&\qquad\frac{1}{(\ell^2+2\ell k+\Delta +i\epsilon)^2}\\
&=\frac{\kappa^2|k^2|}{8\pi^2}\ln\left(\frac{m^2}{m^2+|k^2|}\right),       
\end{split}
\label{yfs1} 
\end{equation}}
where the latter form holds for the UV regime, so that (\ref{indn9}) 
falls faster than any power of $|k^2|$.
With this, taken together with its spinning analog representations, 
for the propagators in the theory, we find that corrections 
such as those illustrated in Fig.~\ref{fig1} 
are UV finite
when superficially they are $D=+4$, 
if $D$ is the standard naive power-counting degree
of divergence. It can be shown that our resummed theory is entirely UV finite~\cite{BW1}. This is consistent with the more phenomenological analyses in Refs.~\cite{laut,reuter2,litim,perc},
which argue for a similar result following Weinberg's asymptotic safety
approach.\par
We note as well that we know of no contradiction between 
our UV analysis and the
important analyses in Refs.~\cite{don1,cav,sola}, which 
deal with the large distance behavior of the theory.\par
\section{Massive Elementary Particles and Black Holes: Final State of Hawking Radiation and Planck Scale Remnants}
As we show in Refs.~\cite{BW1,BW2,BW3,BW4}, when we compute the now UV finite
1-loop contributions to the graviton self-energy, we find the 
improved Newton potential
\begin{equation}
\Phi_{N}(r)= -\frac{G_N M}{r}(1-e^{-ar}),
\label{newtnrn}
\end{equation}
where with 
\begin{equation}
a \cong (\frac{360\pi M_{Pl}^2}{c_{2,eff}})^{\frac{1}{2}}
\end{equation}
we have that~\cite{BW4}
\begin{equation}
a \cong  0.210 M_{Pl}.
\end{equation}
\par
Two consequences of the improved Newton potential are as follows:
\subsection{Elementary Particles and Black Holes}
A massive point particle of rest mass $m$ has its
mass entirely inside of its Schwarzschild radius $r_S=2m/M_{Pl}^2$
so that classically it should be a black hole. We do not expect this
to hold in quantum mechanics.
Focusing on the lapse function
in the metric class 
\begin{equation}
ds^2 = f(r)dt^2-f(r)^{-1}dr^2 - r^2d\Omega^2,
\label{mclass}
\end{equation}
with
\begin{equation}
f(r)=1-\frac{2G(r)m}{r}
\label{lapse}
\end{equation}
and $G(r)$, using (\ref{newtnrn}), given by
\begin{equation}
G(r)=G_N(1-e^{-ar}),
\label{lapse2}
\end{equation}
we see that the Standard Model massive particles all have the 
property that $f(r)$ remains positive as $r$ passes through their
respective Schwarzschild radii and goes to $r=0$ -- the
particle is no longer~\cite{BW2,BW3,BW4} a black hole.
Refs.~\cite{reuter2,maart} have also found that sub-Planck mass black holes
do not exist in quantum field theory.
\par
\subsection{Final State of Hawking Radiation -- Planck Scale Cosmic Rays}
Considering next the evaporation of 
massive black holes, we first note that in Ref.~\cite{reuter2}, 
following Weinberg's~\cite{wein1}
asymptotic safety approach as realized by phenomenological
exact renormalization group methods, it has been shown that the 
attendant running of Newton's constant\footnote{See Ref.~\cite{odint} for a discussion of the gauge invariance issues here.} 
leads to the lapse function
representation, in the metric class in (\ref{mclass}),
\begin{equation}
f(r)=1-\frac{2G(r)M}{r}
\label{massive}
\end{equation}
where $M$ is the mass of the black hole and now
\begin{equation}
G(r)\equiv G_{BR}(r)=\frac{G_Nr^3}{r^3+\tilde{\omega}G_N\left[r+\gamma G_N M\right]}
\label{rnG}
\end{equation}
where $\gamma$ is a phenomenological
parameter~\cite{reuter2} satisfying $0\le\gamma\le\frac{9}{2}$ and
$\tilde{\omega}=\frac{118}{15\pi}$. 
It follows~\cite{reuter2} as well from
(\ref{rnG}) that black holes with
mass less than a critical mass $M_{cr}\sim M_{Pl}$ 
have no horizon. 
Upon joining our result in (\ref{lapse}) onto that in (\ref{rnG})
at the outermost solution, $r_>$, of the equation
\begin{equation}
G_{BR}(r)=G_N(1-e^{-ar}),
\label{match}
\end{equation}
we find the following for the final state of the Hawking process
for an originally very massive black hole: for $r<r_>$, 
we use our result in (\ref{lapse}) for $G(r)$ and for $r>r_>$ we
use $G_{BR}(r)$ for $G(r)$ after the originally massive black hole
has Hawking radiated down to the appropriate scale. 
For the self-consistent 
value $\gamma=0$ and $0.2=\Omega\equiv\frac{\tilde\omega}{G_NM^2}=\frac{\tilde\omega M_{Pl}^2}{M^2}$ for definiteness we find that the 
inner horizon found in Ref.~\cite{reuter2}
moves to negative values of $r$ and 
that the outer horizon moves to $r=0$, so that
the entire mass of the originally very massive black hole radiates away
until a completely accessible Planck scale remnant of mass $M_{cr}'=2.38~M_{Pl}$ is left: It would be expected
to decay into n-body final states, $n=2,3,\cdots$, leading in general
to Planck scale cosmic rays~\cite{BW3,BW4}. 
The data in Ref.~\cite{cosmicray,westerhoff} 
are not inconsistent with this conclusion, which also agrees with
recent results by Hawking~\cite{hawk2}.\par
\section{Conclusions}
In this paper we have presented a new paradigm in the history of 
point particle field theory: a UV finite theory of the union
of quantum mechanics and general
relativity. It holds promise to be a solution 
to most of the outstanding problems
in the union of the ideas of Bohr and Einstein. 
More importantly, it is clear evidence that quantum mechanics,
while not necessarily the ultimate theory, is not an incomplete
theory. This work was supported in part by US DOE grant DE-FG02-05ER41399 and 
by NATO grant PST.CLG.980342. We thank Prof. S. Jadach 
for useful discussions.  

\end{document}